# Schottky-diode design for the world's leading telecommunication

[1]Chi Ho Wong*, [2]Frank Leung Yuk Lam, [2]Xijun. Hu, [1]Anatoly Fedorovich Zatsepin*

[1]Institute of Physics and Technology, Ural Federal University, Yekaterinburg, Russia

[2]Department of Chemical and Biological Engineering, The Hong Kong University of Science and Technology, Hong Kong, China.

**ABSTRACT:** The Schottky diode, BN/GaN layered composite contacting to bulk aluminum, is theoretically plausible to harvest wireless energy above X-band. According to our first principle calculation, the insertion of GaN layers dramatically influences the optical properties of the layered composite. The relative dielectric constant of BN/GaN layered composite as a function of layer-to-layer separation is investigated where the optimized dielectric constant is 3.1. Furthermore, we design another Schottky diode via nanostructuring. Our first principle calculation suggests that the relative dielectric constant of boron nitride monolayer can be minimized to 1.5 only if it is deposited on aluminum monolayer. It is rare to find a semiconductor with the dielectric constant close to 1 which may push the cut-off frequency of Al/BN-based rectenna to the high-band 5G network.

5G wireless communication is based on harvesting wireless energy with help of a local antenna array and low power automated transceiver[1]. The automated transceiver assigned the optimum frequency channels where the local antennas are connected to LTE network. The frequency of the low-band 5G is close to 700 MHz and the transmission of the mid-band 5G is about 2.5-3.7 GHz [2]. The high-band 5G frequencies of 25-40 GHz should provide the fastest internet speed in few gigabit per second [3]. In spite of the fact that the frontier such as Samsung has been investigating the feasibility of mobile communications at frequencies of 28 and 38 GHz[3], the current state-of-the-art Schottky diode harvests wireless energy up to 8GHz -12GHz (X band) only[4]. This X-band Schottky diode is made up of two-dimensional $MoS_2$ sheets and bulk palladium[4] where the relative dielectric constant of $MoS_2$ sheets is above six[5]. Harvesting wireless energy beyond X-band requires the clearance of electric charges in the depletion region rapidly where the cut-off frequency of Schottky diode is inversely proportional to the dielectric constant[6]. High dielectric constant increases the capacitance of Schottky diodes and eventually requires a longer transient time to clear the electric charges[7].

To design a Schottky diode capable of harvesting wireless energy at the high-band 5G signals, the semiconducting side of the Schottky diode needs an ultra-low dielectric constant[7]. It has been an uphill struggle to find semiconductors with the relative dielectric constant close to 1 due to the problem of high electric susceptibilities in semiconductors. For example, the well-known values of dielectric constant of 3D silicon, 3D germanium and 2D $MoS_2$ sheets are as high as 11, 16 and 6, respectively[7,8]. As the dielectric constant depends on size and pressure[9], we decide to modify the nanostructure of Schottky diode and hope for producing ultra-low dielectric constant in order to prepare for 5G world.

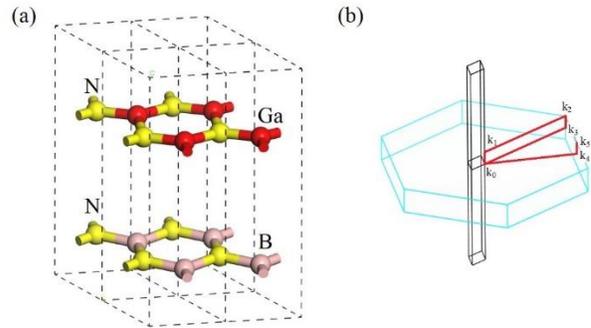

Figure 1: **a** The repeated unit of BN/GaN layered composite. **b** The real (black) and reciprocal (light blue) spaces. The red lines label the direction of k space.

**Methods:**

The BN layers and GaN layers are stacked alternatively as shown in Figure 1. We apply geometric optimization under a boundary condition that the layer-to-layer distance remains at 0.5nm. The geometric optimization is implemented under spin-restricted GGA-PBE functional[20,21]. The convergence tolerance is that the maximum displacement and the maximum force of atom are 0.002 Å and 0.05eV/ Å, respectively. We calculate the band structure and the dielectric constant using the same DFT functional. The maximum SCF cycle is 100 and the SCF tolerance is 2 x 10$^{-6}$eV/atom. The bulk aluminum is attached to BN/GaN layered composite side-by-side. i.e. The Al[100] and BN/GaN [1120] axes are parallel (abbreviated as Schottky diode A).

On the other hands, we examine the semiconducting properties of Schottky diode B in which the boron nitride monolayer is deposited on top of aluminum monolayer. After implementing geometric optimization in the presence of aluminum monolayer using the same DFT functional, we remove all aluminum atoms in the unit cell and proceed to calculate the band gap and the dielectric constant of BN monolayer.

## Results and Discussions:

The band structure of BN/GaN layered composite with the layer-to-layer distance of 0.5nm is plotted in Figure 2. Despite the band gap of the isolated BN sheets is as large as 6eV, the direct band gap of the BN/GaN layered composite is reduced to 1.6eV. On one hand, the insertion of GaN layers strengthens the molecular interaction of the composite laterally[10,11]. On the other hand, the bigger gallium atoms raise the atomic density in each GaN layer so that the composite is subjected to a stronger internal pressure[10,11]. As a result, the wavefunction of electron are overlapped more effectively in the BN/GaN layered composite and a narrower band gap is observed. The theoretical band gaps simulated by the LDA and GGA like PBE, PW91 or BYLP are always underestimated. In principle, the systematic error of bandgap can be corrected by considering a more complex interaction between electrons. The use of the GW corrected GGA or other hybrid functional is a standard approach to tune the theoretical band gap closer to the experimental values. However, there is no universal way to calibrate the ab-initio functional. The Hamiltonian, boundary conditions and simulation parameters of the ab-initio calculation depend on the sample[12-15]. Since the experimental band gap of BN/GaN layered composite is still unknown, we consider the GGA-PBE functional to predict the band gap in an approximated value. Though our theoretical band gap is underestimated, the band gap of BN/GaN layered composite still reaches 1.6eV, which is suitable to a part of Schottky diode.

To identify if the BN/GaN layered composite belongs to p-type or n-type semiconductor, we estimate the the Fermi level ($E_F$) relative to conduction band ($E_c$) and valence band ($E_v$), respectively. Figure 2 and Figure 3 show that the Fermi level of BN/GaN layered composite is located closer to the conduction band and we conclude that the BN/GaN layered composite is a n-type semiconductor. The Fermi level is located nearer to the conduction band because the bigger gallium atom decreases the positive charge density of the composite[16].

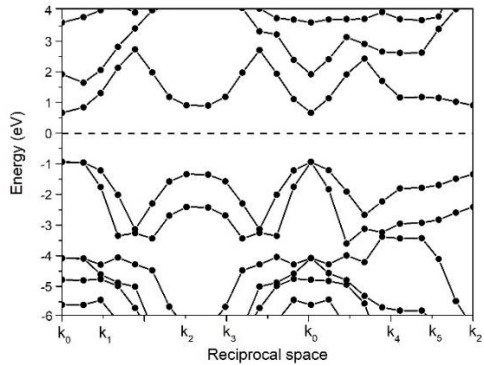

Figure 2: The band structure of BN/GaN layered composite. The layer-to-layer distance is 0.5nm. The Fermi level is shifted to 0eV for convenience.

Figure 3 displays the simplified energy diagrams of Schottky diode A. The work function of aluminum is 4.1eV and the electron affinity of BN/GaN layered composite is 3eV. The Schottky barrier refers to the offset between the work function of metal and the electron affinity of semiconductor[7,16]. The Schottky barrier of most silicon-based diodes is about 0.7eV[17]. By aligning the energy states relative to vacuum level, the Schottky barrier of Schottky diode A is 1.1eV. In other words, the driven voltage of Schottky diode A for forward-bias operation is comparable to the modern silicon-based diodes[7].

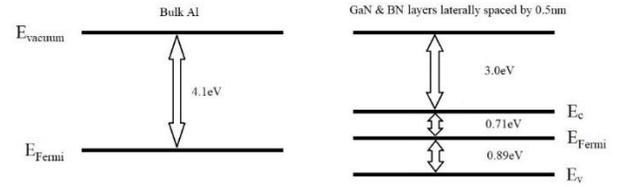

Fig 3: The simplified energy band diagrams of Schottky diode A.

The dielectric function describes the response of the material subjected to a time-dependent electromagnetic field[16]. The excitations induced by the interaction of incident electrons with materials refer to the real and imaginary parts of the dielectric function[18]. We show the relative dielectric constant of BN/GaN layered composite in Table 1. The relative dielectric constant (parallel to the plane) of BN/GaN layered composite is as low as 3.1. In comparison to the modern X-band rectenna consist of two-dimensional $MoS_2$ sheets and palladium[4], the dielectric constant of BN/GaN layered composite is at least 50% lower than the dielectric constant of two-dimensional $MoS_2$ sheets[5]. The use of BN/GaN layered composite as a part of Schottky diode is possible to push the cut-off frequency far beyond X-band because the cut-off frequency is inversely proportional to the capacitance across depletion region. The non-regular atomic layers across the metal-semiconductor interface makes the theoretical prediction of junction resistance inaccurate. However, the electrical resistance of aluminum is ~4 times smaller than that of palladium and hence a low junction resistance is likely expected. The imaginary part of dielectric constant of Schottky diode A and B are close to zero and therefore their phase lags are negligible[18].

Dielectric constant is directly proportional to the permittivity of the medium[16]. If the layer-to-layer distance of BN/GaN layered composite increases to 0.8nm, the relative dielectric constant (parallel to the plane) can reach 2.6 but the band gap raise to 2.5eV. Weakening the layer-to-layer coupling decreases the effectiveness of electric polarization in the medium[18] and presumably drops the permittivity of the substance. The band gap is increased from 1.6eV to 2.5eV because the wider layer-to-layer separation reduces the internal pressure of the composite along the lateral plane[10,16].

Table 1: The dielectric constant of Schottky diode A and B

|  | Materials | Dielectric constant |
|---|---|---|
| Schottky diode A | Bulk Al and BN/GaN layered composite | 3.1 |
| Schottky diode B | Al monolayer and BN monolayer | 1.5 |

While softening the layer-to-layer coupling may yield low dielectric constant in the composite, we proceed to study the optical properties of the isolated BN monolayer. We observe that the relative dielectric constant of the isolated BN monolayer is as low as 2.4. As motivated by this, we decide to examine if the interfacial phonon between aluminum monolayer and boron nitride monolayer can drop the dielectric constant even further. The bond length of the geometrically relaxed aluminum monolayer is increased by 13% when compared to the bulk state. After the BN monolayer is relaxed in the presence of aluminum monolayer, the B-N-B bond angle changes from 120 degree to 116.9 degree and the N-B bond length changes from 1.45Å to 1.48Å. As listed in Table 1, the boron nitride monolayer in Schottky diode B shows the relative dielectric constant of ~1.5 (parallel or perpendicular to the plane) and the simplified energy diagram of Schottky diode B is displayed in Figure 4. The use of Ga monolayer, Pt monolayer, Pd monolayer, Ag monolayer or Au monolayer as the metallic side of the Schottky diode fails to reduce the dielectric constant of BN monolater to ~1.5. In comparison to bulk aluminum, the work function of aluminum monolayer is only 2.5eV because the size-dependent of work function is always expected[19]. Although the energy barrier of Schottky diode B is increased to 1.41V, its ultralow dielectric constant is still capable to harvest wireless energy beyond X-band communication if the driven voltage is above the forward-bias voltage[7]. The dielectric constant of BN monolayer in Schottky diode B is ~4 times smaller than the dielectric constant of two-dimensional $MoS_2$ sheets in the modern X-band Schottky diode[4]. As the cut-off frequency of Schottky diode is inversely proportional to the capacitance across depletion region[7], the ultralow dielectric constant of Schottky diode B may be able to boost the cut-off-frequency towards the high-band 5G signals.

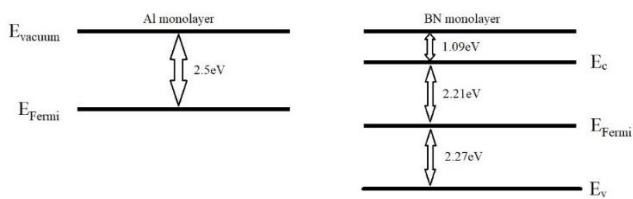

Fig 4. The simplified band diagram of the geometrically relaxed Schottky diode B.

**Conclusion:**

For the purpose of building high-frequency wireless energy harvesting devices, it is plausible to use the BN/GaN layered composite and aluminum to build Schottky diode. We tune the molecular interaction of the interface across the depletion region and find that the dielectric constant of boron nitride monolayer reaches 1.5 which will be further developed to harvest the high-band 5G signals.

**AUTHOR INFORMATION**

Corresponding authors:
ch.kh.vong@urfu.ru, a.f.zatsepin@urfu.ruCorresponding authors:
ch.kh.vong@urfu.ru, a.f.zatsepin@urfu.ru

**Author Contributions**

C.H.Wong and A.F.Zatsepin planed this project. C.H.Wong implemented DFT calculations and wrote manuscript. C.H.Wong and A.F.Zatsepin analyzed data. F.L.Y.Lam and X.Hu proposed the desired Miller indices for the metal-semiconductor interface.

**Funding Sources**

The study was supported by the Ministry of Education and Science of the Russian Federation (Government Task No. 3.1485.2017/4.6) and by Act 211 Government of the Russian Federation (contract no. 02.A03.21.0006).

**Notes**
The authors declare no competing interests.

**ACKNOWLEDGMENT**

We thanks the Ministry of Education and Science of the Russian Federation